\documentclass[11pt]{article}
\usepackage{amssymb,amsmath,amsfonts,mathtools}
\usepackage{commath}
\usepackage{subfigure}
\usepackage{bbm}
\usepackage{epsfig}
\usepackage{wrapfig}
\usepackage{yfonts}
\usepackage{geometry}                		
\usepackage{graphicx}				
\usepackage{color}
\definecolor{hyperref}{RGB}{026,028,185}
\usepackage[bookmarks=true,colorlinks=true,linkcolor=hyperref,citecolor=hyperref,urlcolor=hyperref,bookmarksnumbered]{hyperref}
\usepackage{amsfonts,mathtools}
\usepackage{enumerate}
\usepackage{enumitem}

\usepackage[sort,compress]{cite}
\usepackage[bulletsep]{collref}

\numberwithin{equation}{section}
\def\[{\begin{equation}}
\def\]{\end{equation}}
\newcommand{\be}{\begin{eqnarray}}
\newcommand{\ee}{\end{eqnarray}}

\def\rhoh{\rho_H}
\def\hinf{h_{\infty}}
\def\la{\lambda}

\def\R{\mathcal{R}}

\usepackage{color}

\begin{document}
\renewcommand{\thefootnote}{\arabic{footnote}}
 
\overfullrule=0pt
\parskip=2pt
\parindent=12pt
\headheight=0in \headsep=0in \topmargin=0in \oddsidemargin=0in

\vspace{ -3cm} \thispagestyle{empty} \vspace{-1cm}
\begin{flushright} 
\footnotesize
\end{flushright}%

\begin{center}
\vspace{1.2cm}
{\Large\bf \mathversion{bold}
Non-analyticity of holographic R\'enyi entropy in Lovelock gravity}

 \vspace{0.8cm} {
 V.~Giangreco~M. Puletti \footnote{ {\tt vgmp@hi.is}} ~and~
  R.~Pourhasan \footnote{ {\tt razieh@hi.is}}
}
 \vskip  0.5cm

\small
{\em
University of Iceland,
Science Institute,
Dunhaga 3,  107 Reykjav\'ik, Iceland
}
\normalsize

 \end{center}

\vspace{0.3cm}
\begin{abstract}
We compute holographic R\'enyi entropies for spherical entangling surfaces on the boundary while considering third order Lovelock gravity with negative cosmological constant in the bulk.
Our study shows that third order Lovelock black holes with hyperbolic event horizon are unstable, and at low temperatures those with smaller mass are favoured, giving rise to first order phase transitions in the bulk. 
We determine regions in the Lovelock parameter space in arbitrary dimensions, where bulk phase transitions happen and where boundary causality constraints are met. 
We show that each of these points corresponds to a dual boundary conformal field theory whose R\'enyi entropy exhibits a kink at a certain critical index $n$. 
\end{abstract}
\newpage

\tableofcontents

\setcounter{footnote}{0}
\newpage

\section{Introduction}
\label{sec:intro}

We consider a quantum field theory in a $d$-dimensional Minkowski spacetime when at $t=0$ the system gets separated in two parts, $A$ and its complement $B$, by a $(d-2)$-dimensional hypersurface $\Sigma$. 
A legitimate question to ask is how much the degrees of freedom in the two sub-systems $A$ and $B$ are correlated. Entanglement entropy (EE) and
the R\'{e}nyi entropy (RE) are important measures of this quantum correlation. 
In particular EE across the entangling surface $\Sigma$ is given by
\begin{equation}\label{def_EE}
S(\rho_A)=- \mathrm{tr}\left(\rho_A\ln\rho_A\right)\,,
\end{equation}
where $\rho_A$ is the reduced density matrix of the sub-system $A$, {\it i.e.} the density matrix obtained after integrating out the degrees of freedom in $B$ \cite{Calabrese:2004eu, Calabrese:2005in, Calabrese:2005zw, Calabrese:2009qy}. 
The $n$-th RE, with $n \ge 0$, associated with a quantum system described above is defined as 
\begin{equation}\label{def_RE}
S_n=\frac{1}{1-n}\ln \mathrm{tr}\left(\rho_A^n\right)\,.
\end{equation}
The whole set of eigenvalues of the reduced density matrix $\rho_A$ can be reconstructed by knowing the RE for all the indices $n$. 
For CFT's in flat space, RE exhibits a universal relation to the central charges of the theory, in particular the derivative of RE with respect to $n$ evaluated at $n = 1$ is proportional to the coefficient of the stress tensor two-point function \cite{Perlmutter:2013gua, Dong:2016wcf}. 
Moreover, in the limit where $n\rightarrow1$ RE reduces to EE. 

In general, RE and EE are rather difficult to compute and measure, although remarkable progress in this direction has been made recently \cite{Abanin:2012aa, Daley:2012aa, Islam:2015aa}. 
In quantum field theory RE is mainly computed by means of the so-called replica method \cite{Callan:1994aa, Holzhey:1994we, Calabrese:2004eu, Cardy:2007mb}. 
Here, one replaces the computation of the $n$-th power of the density matrix (and thus the corresponding partition function) with that of the density matrix of a theory which consists of $n$ copies of the original quantum field theory. 
This amounts to computing the Euclidean partition function on a geometry with conical singularity. 
%
Although a direct ``holographic translation'' of the replica approach might involve conically singular geometries, which are generally
difficult to deal with and may not lead to the correct results \cite{Fursaev:2006ih, Headrick:2010zt}, RE can be studied holographically.   

In holographic theories EE can be computed by the Ryu-Takayanagi (RT) formula%
\footnote{The formula has recently  been proved in \cite{Lewkowycz:2013nqa, Dong:2016hjy}, a first attempt to prove it was presented in \cite{Fursaev:2006ih}, cf. \cite{Rangamani:2016aa} for a recent review on holographic EE.} 
involving minimal surfaces which extend into the bulk and end on the boundary entangling surface \cite{Ryu:2006bv, Ryu:2006ef, Hubeny:2007xt}. 
A somewhat similar prescription for RE has been provided only recently in \cite{Dong:2016fnf}: RE can be determined by computing the area of cosmic branes which back-react with the bulk geometry. Despite its beauty and geometric foundation, the prescription in \cite{Dong:2016fnf} can be arduous to handle in general cases. 

Another remarkable approach to compute holographic EE for spherical entangling surfaces is the one proposed by Casini, Huerta, and Myers (CHM) \cite{Casini:2011kv}, extended to the holographic RE in \cite{Hung:2011nu}. 
It consists of a conformal mapping on the CFT which takes us from an Euclidean conically singular geometry to an Euclidean smooth thermal hyperboloid.
The gravity dual of such a thermal CFT (if it exists) is a black hole with hyperbolic event horizon in asymptotically AdS (AAdS) spacetimes.  
Hence, the CHM map relates the RE of the original CFT to the free energy of  AAdS hyperbolic black holes. 
The index of the RE is translated into the inverse of the black hole temperature (compared to some reference temperature).
Therefore, the knowledge of RE at any $n$ (quantum entanglement spectrum) requires the knowledge of free energy (and thus thermal entropy) of a hyperbolic black hole in AdS at any temperature. We will review the crucial  steps of the CHM map in section \ref{sec:RE-HRE}.

The advantages of CHM approach are twofold. 
First of all, it avoids conical singularities and related problems \cite{Fursaev:2006ih, Headrick:2010aa}, by working on a thermal ensemble which makes the boundary geometry perfectly smooth and straightforwardly treatable via standard holographic techniques.
Second, it applies to any gravity theory (assuming they have a CFT dual) and in particular to higher derivative gravities \cite{Hung:2011nu, Hung:2014npa, Chu:2016tps, Bianchi:2016xvf, Camps:2016gfs}, unlike the RT formula which needs to be corrected \cite{Hung:2011aa, deBoer:2011wk, Myers:2010aa, Myers:2010ab, Ogawa:2011aa, Dong:2013aa, Camps:2013aa}. 

In this manuscript we apply the CHM approach to study RE of holographic CFT, dual to higher derivative gravity theories, in particular the so-called third order Lovelock gravity \cite{Lovelock:1971yv, ref1}, in an asymptotically AdS spacetime. 
Lovelock gravities are interesting generalizations of Einstein gravity, which are ghost-free and 
living in dimensions (strictly) greater than four with small coupling constants, {\it i.e.} small corrections to general relativity. 
In third order Lovelock gravity the Einstein-Hilbert action is corrected with terms proportional to $\R^2$ (with $\R$ the curvature scalar), also known as Gauss-Bonnet gravity%
\footnote{For the relation between Gauss-Bonnet gravity and string theory see for example \cite{Zwiebach:1985uq, Metsaev:1987zx}.}, 
and $\R^3$ with dimensionless coupling constants $\lambda$ and $\mu$, respectively. 
We will review basic aspects of Lovelock gravity in section \ref{sec:Lovelock-gravity}. 
These theories have proven useful in exploring various properties of holographic theories, as for example the viscosity bounds \cite{Brigante:2007nu, deBoer:2009pn, Camanho:2010aa}, although at intermediate energy scales they might become problematic \cite{Camanho:2014aa}.%
\footnote{``Intermediate energy scales'' is referred to energy scales where higher derivative corrections are important but the theory is still weakly coupled \cite{Camanho:2014aa}.}
However, in this work we always assume that Lovelock couplings are small positive numbers, satisfying  constraints coming from boundary causality \cite{Camanho:2009hu}, as will be reviewed in section \ref{sec:CausCos}. 
We work in a classical regime, therefore, the necessary additional degrees of freedom mentioned in  \cite{Camanho:2014aa} to cure causality are not relevant to our discussions.

As we will see in section \ref{sec:phase-trans} third order Lovelock theories reveal interesting and unusual features, not present in Einstein gravity and not even in Gauss-Bonnet gravity.  
Our study unveils first order phase transitions between coexisting hyperbolic black holes in third order Lovelock gravity.  
Such phase transitions were originally observed in a study by one of the authors in \cite{Dehghani:2009zzb}, however, only for the special case where $\mu=\la^2/3$. 
In the present work, by exploring the complete Lovelock parameter space spanned by $\{\lambda, \mu\}$, we find that in any given dimension, there are regions where the hyperbolic black holes with smaller mass are favoured at low temperature. 
Some of these phase transitions are in the range of parameter space which is excluded by boundary causality, including the specific case $\mu=\la^2/3$. Nevertheless we can still observe plenty of phase transitions in the causal regions. 
Figure \ref{fig:mulambda678} shows an example of regions where causality and bulk phase transitions overlap in 7, 8 and 9-dimensions. 
For instance, in seven dimensions ($d=6$) we observe phase transitions in the causal range for $0.25 \le \lambda \le 0.387$ and $0.024 \le \mu \le 0.105$, or in terms of actual Lovelock couplings, for $0.021 \le {\alpha_2\over L^2} \le 0.032$ and $0.001\le {\alpha_3 \over L^4} \le 0.004$. Note that, in the causal range where phase transitions happen the Lovelock couplings in  (\ref{3OL}) are still small enough that $\mathcal{L}_2$ and $\mathcal{L}_3$ can be considered as perturbations to the Einstein term.
 
Furthermore, the first order nature of the phase transitions indicates a discontinuity in the black hole thermal entropy. 
As mentioned earlier in CHM holographic approach, the RE of a boundary CFT is related to the free energy, and thus to the 
thermal entropy of black holes. 
It is then interesting to investigate the effects of these bulk phase transitions on the boundary field theory RE.

This is the main focus of this work and the results are discussed in section \ref{sec:RE-Lovelock}. 
The holographic RE for third order Lovelock gravity was already computed in \cite{Hung:2011nu, Hung:2014npa}. 
The novelty here is to take into account that such black holes undergo phase transitions, to systematically span the {\it causal} parameter space, and analyse the consequences for the dual RE. 
Connections between RE and bulk phase transitions have been previously studied in \cite{Belin:2013dva, Belin:2014mva}. 
However, there are two main differences here. 
First of all, our system is purely gravitational, dual to a CFT in its vacuum state with the only corrections coming from the corrections of the coupling constants. 
In \cite{Belin:2013dva}  the authors holographically  computed RE by considering Einstein gravity with the addition of a scalar field (similarly in \cite{Belin:2014mva} for the case of a charged system), and the instability of hyperbolic black holes is due to the development of hair. 
Second, our phase transition is first order, while in \cite{Belin:2013dva, Belin:2014mva} it is second order. 
This has a crucial effect on the RE: our results show that for strongly coupled dual CFT's  the RE displays a kink at a critical index $n$ which results in the non-analyticity of RE with respect to $n$ nearby the kink.

While our findings are particularly interesting for $d=6$ where we have known examples of AdS/CFT dualities, they are valid as well for $d$-dimensional field theories with $d > 6$. 
In fact,  
from the bulk point of view the number of dimensions $D=d+1$ is a mere parameter, and it is interesting to explore its effect on the system.
Our analysis shows that $D=7$ is not special: in any dimension $D=d+1\ge 7$ it is possible to find regions of the parameter space where black holes are unstable and where the would-be boundary field theory is causal (even though these regions shrink as we increase the number of spacetime dimensions). 
Existence of higher ($d>6$) dimensional CFT's is still an open question, {\it e.g.} \cite{Witten:2007ct}. However, assuming that a dual CFT exists, here third order Lovelock theories can serve as a toy-model: they allow us to straightforwardly  carry on computations, and thus, to explore the role of higher derivative gravity in this context. 
This kind of approach has turned out to be helpful in the past, {\it e.g.} cf.  \cite{Brigante:2007nu, deBoer:2009pn, Camanho:2010aa} on the discussion of  the viscosity bound or \cite{Myers:2010aa, Myers:2010ab}  for the discovery of the F-theorem.
For this reason, we hope that the holographic system studied in this work might be instructive to predict novel features for strongly coupled higher dimensional conformal field theories.


\section{Holographic R\'enyi entropy}
\label{sec:RE-HRE}


We will be interested in thermal states, so it is useful to understand the role of R\'enyi entropies in this case. 
A description of the quantum R\'enyi entropy for a thermal state in terms of the free energy has been discussed in \cite{Baez:2011aa}. 
Suppose we have a physical system which is in thermal equilibrium at temperature $T_0$. 
When the system is ``quenched'' and the temperature is lowered by a factor $n$, the R\'enyi entropy is a measure of the maximum amount of work (divided by the difference of temperature) the system can do in reaching the new equilibrium state and is given by
\begin{equation}\label{SnFT}
S_n(T_0)=-\frac{F(T)-F(T_0)}{T-T_0}\,,
\end{equation}
where
\begin{equation}
n=\frac{T_0}{T}\,.
\end{equation}
In the limit $n\rightarrow1$ the right hand side of expression \eqref{SnFT} gives the usual relation for thermal entropy, {\it i.e.}
\begin{equation}\label{Sthe}
S_{thermal}(T_0)=-\frac{dF}{dT}\mid_{T=T_0}\,,
\end{equation}
which can be then used to rewrite the R\'enyi entropy in (\ref{SnFT}) as
\begin{equation}
S_n(T_0)=\frac{n}{n-1}\frac{1}{T_0}\int^{T_0}_{T_0/n} S_{thermal}(T')dT'\,.\label{RenyiEnt}
\end{equation}
%

\vskip 0.5 cm
We now review the main steps of the CHM approach to compute holographic R\'enyi entropy~\cite{Casini:2011kv, Hung:2011nu}. 
Let us start with a CFT in $\mathbb{R}^{1,d-1}$ in the vacuum state. The system is at zero temperature, and we introduce a $(d-2)$-dimensional {\it spherical} entangling surface $\Sigma$. 
The conformal transformations found in \cite{Casini:2011kv} map the reduced density matrix of a CFT in flat spacetime  to a {\it thermal} density matrix of a CFT on a hyperbolic geometry $\mathcal H\equiv \mathbb{R}\times \mathbb{H}^{d-1}$,
 where $\mathbb{H}^{d-1}$ is a hyperbolic $(d-1)$-dimensional space.
 The radius of the curvature of the hyperbolic plane  matches the radius $R$ of the entangling surface $\Sigma$, and in particular the temperature is given by the inverse of $R$.%
\footnote{The conformal transformations found in \cite{Casini:2011kv} map the causal development of the region inside $\Sigma$ to a Rindler wedge, which is in turn mapped to a hyperbolic plane $\mathcal H\equiv \mathbb{R}\times \mathbb{H}^{d-1}$. The crucial point is that the vacuum state of the original CFT is mapped to a state in $\mathcal H$ which looks thermal with respect to the Hamiltonian generating the time evolution in $\mathcal H$ (we refer the reader to the original reference \cite{Casini:2011kv} for more details), hence the relation among the density matrices.}
The mapping among density matrices extends to the entropy. Hence, the entanglement entropy of a spherical entangling $(d-2)$-dimensional surface of radius $R$ in a CFT  in flat spacetime is equivalent to the thermal entropy of a CFT at temperature $T_0=1/2\pi R$ in  a hyperbolic geometry $\mathbb{R}\times \mathbb{H}^{d-1}$.

According to the AdS/CFT correspondence, a thermal state in the boundary CFT is dual to a black hole in the bulk geometry. 
Since the CFT has been defined on a hyperbolic plane, by matching the geometry on both sides of the duality, the appropriate black hole to consider in the bulk is the so-called topological black hole, {\it i.e.} one with hyperbolic horizon~\cite{Casini:2011kv}.
The Hawking temperature of the black hole is then related to the temperature of the dual field theory according to the usual AdS/CFT dictionary. 
Therefore, in this framework, the entanglement entropy across $\Sigma$ is given by the horizon entropy of a hyperbolic AdS black hole \cite{Casini:2011kv}. 

The procedure described above can be extended in a straightforward manner to the holographic calculation of R\'enyi entropies for a spherical entangling surface  \cite{Hung:2011nu, Hung:2014npa}. We have seen above that the computation of R\'enyi entropies requires the knowledge of the system at a temperature $T$ given by $T_0/n$, see for example \eqref{SnFT}. 
Holographically, this means that we need to extend the AdS hyperbolic black hole solution to any $T=T_0/n$.

\section{Thermodynamics of Lovelock black holes}
\label{sec:Lovelock-BH}

In section  \ref{sec:Lovelock-gravity} we recall some basic features of third order Lovelock gravity with a negative cosmological constant and the corresponding hyperbolic black hole solutions. 
In section \ref{sec:CausCos} we review the constraints on the Lovelock coupling constants $\{ \lambda, \mu\}$ imposed by requiring that the boundary CFT is causal. In section \ref{sec:phase-trans} we study the thermodynamics properties of these black holes as a function of the couplings in arbitrary dimensions.

\subsection{Topological Lovelock black holes}
\label{sec:Lovelock-gravity}

In a spacetime with dimensions higher than four, Einstein gravity is not the most
general gravitational theory sharing the basic properties of standard general relativity, that is field equations are generally covariant and contain at most second order derivatives of the
metric. 
Based on these assumptions, the action for the most general gravity theory in $(d+1)$-dimensions is written as Lovelock gravity with the Lagrangian in the form
\cite{Lovelock:1971yv, ref1}
\be\nonumber
\mathcal{L}= \sum\limits_{p=1}^{[d/2]}\, \alpha_p\,\mathcal{L}_p\,,
\ee
where $\mathcal{L}_1$ is the Einstein-Hilbert term, $\mathcal{L}_2$ is the Gauss-Bonnet term, $\mathcal{L}_3$ is a third order Lovelock term, and so on. 
Here, we consider up to third order Lovelock gravity with a negative cosmological constant, therefore we restrict ourselves to the following action
\begin{equation}
\label{3OL}
I=\frac{1}{2 \ell_p^{d-1}}\int d^{d+1}x \sqrt{-g} \left(\frac{d(d-1)}{L^2} + \mathcal{R} + \alpha_2\mathcal{L}_2 + \alpha_3\mathcal{L}_3\right)\,,
\end{equation}
where $\R$ is the curvature scalar in the bulk, and
\begin{eqnarray}
&&\mathcal{L}_2=\R_{ijkl} \R^{ijkl}-4\R_{ij}\R^{ij}+\R^2,\\
&&\nonumber\\
&&\mathcal{L}_3=2\,\R^{ijkl}\R_{klmn}\R^{mn}_{\,\,ij}+8\,\R^{ij}_{km}\R^{kl}_{jn}\R^{mn}_{il}+24\,\R^{ijkl}\R_{kljm}\R^{m}_{i}\nonumber\\
&&\hspace{.6cm}+\,3\R\,\R^{ijkl}\,\R_{klij}+24\,\R^{ikjl}\R_{ji}\R_{lk}+16\,\R^{ij}\,\R_{jk}\,\R^{k}_{\,i}-12\,\R\,\R^{ij}\,\R_{ji}+\R^3\,.
\end{eqnarray} 
$L$ is the scale of the cosmological constant described by the first term in \eqref{3OL}, $\ell_p$ is the Plank length, $\alpha_2$ and $\alpha_3$ are the second and third order Lovelock couplings with dimensions of a $(\text{length})^2$ and $(\text{length})^4$, respectively%
\footnote{We normalize the action \eqref{3OL} such that $\alpha_1=1$.}.
$\mathcal L_2$ and $\mathcal L_3$ are not zero only for dimensions strictly higher than four and six, respectively. They are simply proportional to the corresponding Euler density in four and six dimensions.
For convenience, the Lovelock coefficients are written in terms of dimensionless parameters as follows
\begin{equation}
\alpha_2=\frac{L^2\lambda}{(d-2)(d-3)}\,,\qquad\alpha_3=\frac{L^4\mu}{(d-2)(d-3)(d-4)(d-5)}\,.
\end{equation} 
Here $\lambda$ and $\mu$ are chosen to be positive.

By varying the action (\ref{3OL}), one obtains the equations of motion up to third order in Lovelock coefficients as follows
\begin{equation}
G_{ij}-\frac{d(d-1)}{2 L^2} g_{ij}+ \frac{L^2\lambda}{(d-2)(d-3)} G_{ij}^{(2)}+\frac{L^4\mu}{(d-2)(d-3)(d-4)(d-5)} G_{ij}^{(3)}=0\,,\label{3Leom}
\end{equation}
where $G_{ij}=\R_{ij} - \frac{1}{2} g_{ij} \R$ is the Einstein tensor and 
\begin{eqnarray}
&& G_{ij}^{(2)} = 2 \left( \R_{iklm} \R_{j}^{\,\,\,klm} - 2 \R_{ik} \R_{j}^{\,\,\,k} - 2 \R_{ikjl} \R^{kl}+ \R \R_{ij}\right) - \frac{1}{2} g_{ij} \mathcal{L}_2\,,\\
&& G_{\mu\nu}^{(3)} = 3 \big( \R_{ij} \R^2 - 4 \R_{ij} \R^{kl} \R_{kl} + \R_{ij} \R_{klmn} \R^{klmn} - 4 \R_{ikjl} \R^{kl} \R \nonumber\\
&&\hspace{1cm}+\, 8 \R_{ikjl} \R^{kmln} \R_{mn} + 8 \R_{ikjl} \R^{km} \R^{\,\,\,l}_{m} - 4 \R_{ikjl} \R^{k}_{\,\,\,mnp} \R^{lmnp} - 4 \R_{ik} \R^{k}_{\,\,\,j} \R \nonumber\\
&&\hspace{1cm} +\, 8 \R_{iklm} \R^{l}_{\,\,j} \R^{km} + 4 \R_{iklm} \R^{lmk}_{\hspace{.46cm}n} \R^{n}_{\,\,\,j} + 2  \R_{iklm} \R^{\,\,klm}_{j} \R- 4 \R_{iklm} \R^{lm}_{\hspace{.32cm}jn} \R^{kn}\nonumber\\
&&\hspace{1cm} +\, 4 \R_{jklm} \R^{lmkn} \R_{in} + 2 \R_{iklm} \R^{knp}_{j} \R^{\,\,\,\,lm}_{np} + 8 \R_{ik} \R_{jl} \R^{kl} - 8 \R_{iklm} \R_{j\,\,\,\,\,n}^{\,\,kl} \R^{mn} \nonumber\\
&&\hspace{1cm}+ 8 \R_{jklm} \R^{l}_{\,\,i} \R^{km} - 8 \R_{iklm} \R^{ln}_{\,\,\,\,jp} \R^{mpk}_{\hspace{.46cm}n}\big) - \frac{1}{2} g_{ij} \mathcal{L}_3\,.
\end{eqnarray}

We will consider spherically symmetric hyperbolic black holes, thus we can employ the following metric ansatz 
\be
ds^2= -\left(-1+h(\rho) {\rho^2 }\right) N^2 dt^2+ L^2\left({d\rho^2\over \left(-1+h(\rho) {\rho^2}\right)}  +\rho^2 d\Sigma_{-1,d-1}^2\right)\,,\label{metric}
\ee
where $d\Sigma_{-1,d-1}$ is the metric of a  $(d-1)$-dimensional unit hyperboloid  and $N$ is a constant introduced to have a convenient normalization of the time coordinate. 
Clearly, $h(\rho)$ has to be a solution of the equations of motion \eqref{3Leom}.
Plugging the ansatz \eqref{metric}  into the equations (\ref{3Leom}), we obtain a simple expression for the integral of motion
\be
\label{master_eq}
\rho^d \left(1-h(\rho)+\lambda\, h(\rho)^2 -\mu \, h(\rho)^3\right)=\text{const} \equiv  m \,.
\ee
Note that $m$ is the {\it dimensionless} black hole conserved charge, therefore a measure of its mass.
$m$ \eqref{master_eq} can be expressed in terms of the  (dimensionless) black hole horizon $\rhoh$, defined by $g_{tt}(\rhoh)=0$, that is
\be\label{mass}
m=\rhoh^{d-6}\left(\rhoh^6-\rhoh^4+\la\rhoh^2-\mu\right)\,.
\ee
Similarly,  the integral of motion (\ref{master_eq}) evaluated at the boundary $\rho\rightarrow\infty$ defines the asymptotic value of $h(\rho)$,{\it i.e.} $\hinf$, as
\be\label{def_hinf}
1-\hinf+\lambda \, \hinf^2 -\mu\,  \hinf^3=0\,. 
\ee
A convenient choice for the normalization constant $N$ is \cite{Hung:2011nu}
\be
N^2= {L^2 \over \hinf \, R^2}\,,
\ee
in this way the curvature scale of the hyperbolic spatial slices is $R$ in the boundary CFT on $\mathbb{R}\times \mathbb{H}^{d-1}$. 

The metric (\ref{metric}) asymptotically represents a pure AdS spacetime with radius $\widetilde{L}$ where
\be\label{def_Ltilde}
\widetilde{L}^2=\frac{L^2}{\hinf}\,,
\ee
or in other words, the effective cosmological constant is, in fact, 
\begin{equation}
\Lambda_{\text{eff}}=\frac{1}{\widetilde{L}^2}=\frac{\hinf}{L^2}\,.
\end{equation}
In principle, equation (\ref{def_hinf}) could have three real distinct solutions provided the discriminant is positive. Therefore there exist three different effective cosmological constants. However, if the discriminant of (\ref{def_hinf}) vanishes, all three solutions coincide. This happens at $\la=1/3$ and $\mu=1/27$, thus the theory has maximum degeneracy and the full symmetry of AdS is recovered for this particular choice of Lovelock parameters.  

By examining the equations of motion (\ref{master_eq}), it is straightforward to find that there is always a {\it unique} solution for $h(\rho)$ which is {\it real} everywhere provided that, in any given dimension, the Lovelock coefficients satisfy the following condition
\be
\mu \ge \frac{\la^2}{3}\,.\label{mulambda}
\ee
From this point forward parameters are chosen such that the condition (\ref{mulambda}) holds. Also the discriminant of (\ref{def_hinf}) is strictly negative when the inequality in  (\ref{mulambda}) is fulfilled and therefore a fixed $\{\la, \mu\}$ results in only one effective cosmological constant, {\it i.e.} a unique AdS at the boundary. 

The metric solution for generic $\la$ and $\mu$ is easily obtained from equation (\ref{master_eq}) as
\footnote{We partly borrow notation used in \cite{Dehghani:2009zzb}. }%
%
%
\be\label{h_general}
h(\rho) &=& {\la\over 3\mu}\left[1+ \left(\sqrt{\Gamma+J(\rho)^2}+J(\rho)\right)^{1/3}-\left(\sqrt{\Gamma+J(\rho)^2}-J(\rho)\right)^{1/3}\right]\,, ~~~\\ \nonumber
J(\rho)&\equiv& 1- {9 \mu\over 2\la^2}+{27 \mu^2\over 2\la^3} K(\rho)\,, \qquad K(\rho)\equiv 1- { m \over \rho^d}\,, \qquad \Gamma \equiv \left({3\mu\over \la^2}-1\right)^3\,.
\ee
In the following we also express other thermodynamic formulae which will be used to aid in further calculations.%
\footnote{We do not follow the conventions adopted in \cite{Hung:2014npa, Hung:2011nu}, however our results if written in terms of the parameter $x$, {\it i.e.}
$
 x= {\rhoh \, \sqrt{\hinf}}\,,
$
agree exactly with theirs.}
One can assume a black hole as a thermodynamic system \cite{Hawking:1971tu, Bekenstein:1973ur, Bekenstein:1972tm, Bekenstein:1974ax} and define the Hawking temperature \cite{Hawking:1974rv, Hawking:1974sw} as
\be
\label{temp}
T &=& {N \over 4\pi L } \left| \partial_\rho g_{tt}(\rhoh)\right|
= \frac{d \rho _H^6-(d-2) \rho _H^4+(d-4) \lambda  \rho _H^2-(d-6) \mu }{4 \pi  R\,  \sqrt{h_{\infty }}\,
   \rho _H \left(\rho _H^4-2 \lambda  \rho _H^2+3 \mu \right)}\,.
\ee
With our conventions, the AdS solution corresponds to a temperature given by 
\be\label{ads-temp}
T_0={1\over 2 \pi R}\,.
\ee
This can be seen by using the relation \eqref{mass} with $m=0$, and recalling that for an AdS spacetime the function $h(\rho)$ is the constant $\hinf$, which implies $\rhoh={1\over\sqrt \hinf}$. 

The ADM mass can be worked out in a straightforward manner from $m$ \eqref{mass}, and it is given by
\be
\label{ADM_mass_general}
M &=&
V_{\Sigma } \, \left(\frac{L}{\ell_p}\right)^{d-1}\, \frac{(d-1) \rho _H^{d-6}  \left(\rho _H^6-\rho _H^4+\lambda  \rho
   _H^2-\mu \right)}{2 \,R\, \sqrt{h_{\infty }}}\,,
\ee
where $V_\Sigma$ is the volume of the hyperboloid $\Sigma_{-1,d-1}$. 
The horizon entropy can be computed from the Wald entropy formula \cite{Wald:1993nt, Iyer:1994ys, Jacobson:1993vj}, and it results in \cite{Hung:2014npa}
\be
\label{entropy_general_d}
S&=& 
2\, \pi\,  V_{\Sigma }\left(\frac{L}{\ell_p}\right)^{d-1} \left( \rhoh^{d-1} +3 \mu {d-1\over d-5}\rhoh^{d-5}-2\la {d-1\over d-3}\rhoh^{d-3}\right) \,.
\ee
$V_\Sigma$ is a divergent quantity, and in particular its  leading behaviour is proportional to ${ \epsilon^{2-d}}$, where $\epsilon$ is a short-distance cut-off \cite{Casini:2011kv, Hung:2011nu}. Such UV-divergences are expected, and they correspond to the (divergent) terms responsible for the so-called area law in the boundary field theory. 

In a classical regime the bulk partition function reduces to the exponential of (minus) the regularized classical on-shell action $S_{E, \text{reg}}$, thus the black hole free energy is simply given by
\be
F= T \ S_{E, \text{reg}}\,. 
\ee
$S_{E, \text{reg}}$ can be computed by extending holographic counter-term methods to general Lovelock theories, explicitly developed in~\cite{Yale:2011dq, Mehdizadeh:2015cya} and initiated in \cite{Dehghani:2006ws, Astefanesei:2008wz, Cvetic:2001bk}.\footnote{For an alternative regularization approach to derive similar counter-terms, we refer an interested reader to \cite{Kofinas:2007ns}.} 
The final result can be written as
\be\label{def_free_en}
F= E_0+ M- TS\,,
\ee
where $M$, $T$, and $S$ are given by \eqref{ADM_mass_general}, \eqref{temp}, and \eqref{entropy_general_d} respectively. 
$E_0$ is a finite constant term which arises from the counter-term methods, and accounts for the Casimir energy.
It depends upon the Lovelock couplings $\{\lambda, \mu\}$, but not on the horizon data.%
\footnote{In particular when $\lambda$ and $\mu$ are set to zero $E_0$ reduces to the hyperbolic AdS Casimir energy \cite{Emparan:1999pm, Emparan:1999gf, Vanzo:1997gw, Kraus:1999di}.}
Consequently, it leads to an overall shift in the free energy. 

Since we will be interested in comparing free energies of coexisting black hole solutions at any given $\{ \lambda, \mu\}$ and $d$ (section \ref{sec:phase-trans}), the Casimir energy $E_0$ will not play any role, and we can safely work with the following free energy density per unit volume
\be\label{curlyF}
\mathcal F= { F- E_0 \over V_\Sigma \left(\frac{L}{\ell_p}\right)^{d-1} T_0}\,.
\ee
Here, we introduce the thermal entropy density which will be useful later, as
\be\label{curlyS}
\mathcal S={S\over V_{\Sigma }\left(\frac{L}{\ell_p}\right)^{d-1}}\,,
\ee
where $S$ is the thermal entropy \eqref{entropy_general_d}. 
Note that $\mathcal F$ and $\mathcal S$ are dimensionless. 

In the rest of this section we will investigate the thermodynamics of hyperbolic  black holes in the full parameter space $\{ \la\,, \mu\}$ of third order Lovelock gravity in arbitrary dimensions. In particular, we find that in any given dimension for certain values of $\{ \la\,, \mu\}$ there exist multiple isothermal black holes, a fact that is a signal of a possible phase transition in the theory. However, before moving to identify where in the parameter space the phase transition will occur, crucial limits on $\{ \la\,, \mu\}$ should be taken into account which arise from the causality constraints of the CFT boundary theory.
These constraints will be briefly discussed in the following section \ref{sec:CausCos}.


\subsection{Causality constraints on the Lovelock parameters}
\label{sec:CausCos}
\begin{figure}[t!]
\begin{center}
    \includegraphics[width=.6\textwidth]{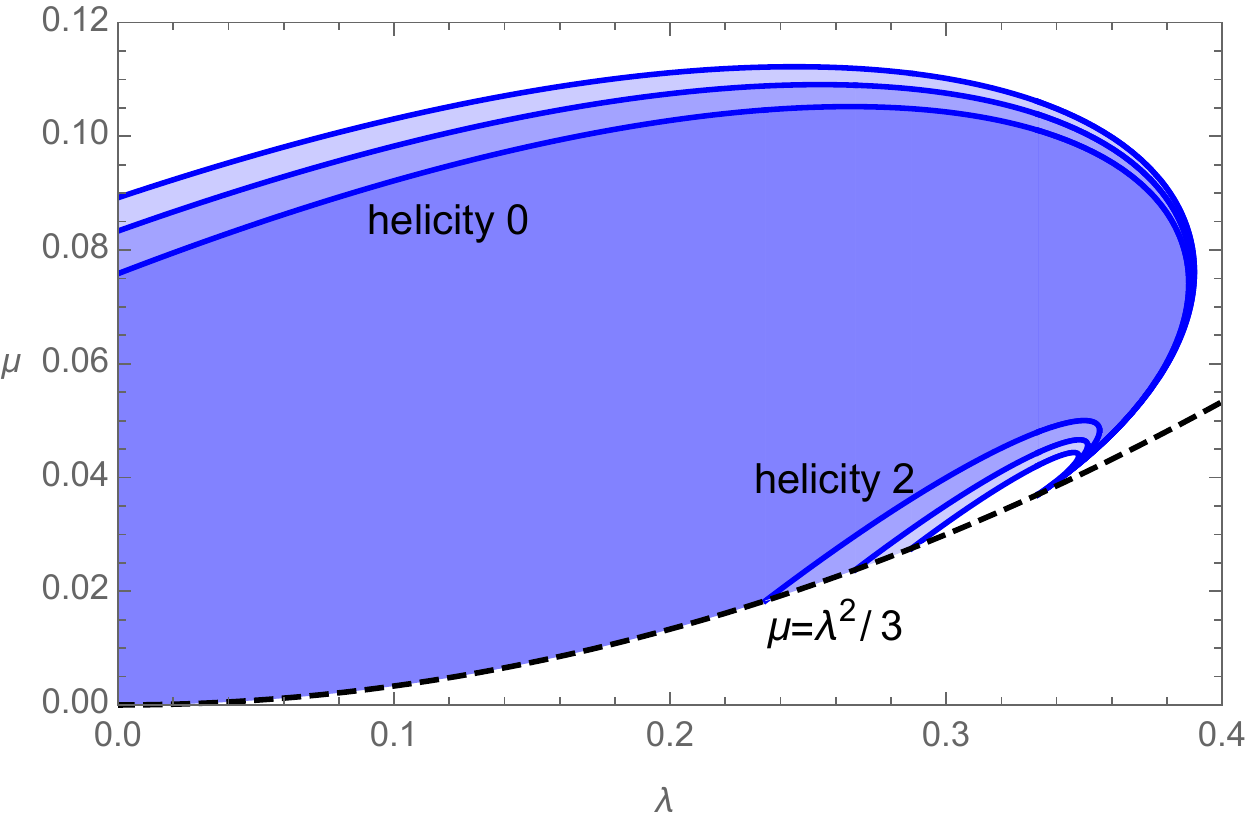}
    \caption{Shadowed blue indicates the allowed regions by causality for Lovelock parameters in 7, 8 and 9-dimensions which grow as dimensionality increases.\label{fig:Causd678} }
\end{center}
\end{figure}

Demanding causality of the boundary theory, the fact that the velocity of any signal propagating on the boundary should not exceed the speed of light, will introduce constraints on the Lovelock parameters. These constraints have been well studied in the literature for Gauss-Bonnet \cite{Brigante:2008aa, Buchel:2009aa, Buchel:2009ab, Camanho:2009aa, Hofman:2009aa, deBoer:2009pn} and third order Lovelock gravities \cite{Camanho:2009hu, deBoer:2009gx, Camanho:2010aa}. 
Here, we follow the results obtained in \cite{Camanho:2009hu} for third order Lovelock gravity where the constraints have been derived using the perturbations of metric as well as shock waves calculation. While we encourage an interested reader to find the details of calculations in \cite{Camanho:2009hu} and reference therein, we only express the final results here. In general, there exist three modes propagating on the boundary: helicity 2, helicity 1 and helicity 0 gravitons. The requirement that each mode propagates with the velocity lower than the speed of light imposes the following constraints:
\begin{eqnarray}\label{CausCons}
\text{helicity 2}&:& \,\,1 - \frac{2 \left(d^2 - 5 d+10\right)  }{(d-4) (d-3)} \la \, h_{\infty} + \frac{3 \left(d^2 - 3 d+8\right) }{(d-4) (d-3)}\mu \, h_{\infty}^2 \ge 0\,,\nonumber\\
\text{helicity 1}&:& \,\,1 + \frac{4}{(d-3)}  \la \, h_{\infty}  - \frac{3 (d+1)}{(d-3)} \mu \, h_{\infty}^2 \ge 0\,,\\
\text{helicity 0}&:& \,\,  1 + \frac{2 (d+1) }{(d-3)} \la \, h_{\infty} - \frac{3 (3 d-1) }{(d-3)} \mu \, h_{\infty}^2 \ge 0\,,\nonumber
\end{eqnarray}
where $h_{\infty}$ is governed by equation (\ref{def_hinf}). 

Exploring the space of Lovelock parameters while respecting constraints (\ref{CausCons}), one finds that the causality of helicity 2 boundary gravitons will set a lower bound on the parameters while the causality of the other two modes imposes an upper bound on the allowed region of $\{\la, \mu\}$. 
However, helicity 0 constraint is always more stringent than helicity 1. 
Therefore, at the end, the allowed region due to causality is identified by the helicity 2 and helicity 0 modes. This is true in any arbitrary dimension. 

The shadowed blue region in figure \ref{fig:Causd678} shows the region in the parameter space fulfilling the causality constraints (\ref{CausCons}), as well as reality constraints, in 7, 8 and 9-dimensions, respectively. 
Recall that in any given dimension, the full metric solution \eqref{metric} is real everywhere whenever the condition \eqref{mulambda} holds, {\it i.e.} $\mu\ge \la^2/3$. 
Therefore any region below the parabola $\mu=\frac{\la^2}{3}$ is excluded although it may be allowed by boundary causality.  
Figure  \ref{fig:Causd678} clearly indicates that the allowed region grows as we move to higher dimensions. 


\subsection{Phase transitions for $\mu\ge \la^2/3$}
\label{sec:phase-trans}

This section is devoted to determine where in the parameter space $\{\la, \mu\}$ of third order Lovelock gravity, phase transitions happen for hyperbolic black holes in arbitrary dimensions. 
In order to identify whether a thermal phase transition would occur, we need to look for the existence of isothermal black hole solutions. 
To do so, one should examine the behaviour of the temperature as a function of black hole mass, {\it i.e.} $T(M)$. 
If we find that temperature is a non-monotonic function of mass, then at a given temperature there are coexisting black hole solutions with different masses, or different horizon radii, which signals the possibility of a thermal phase transition in the gravitational system. 
In order to confirm that a phase transition happens, one should further compare the free energy of isothermal solutions. 
The way to investigate non-monotonicity of $T(M)$ is to examine whether ${dT\over dM}$ has a real solution or not: if the derivative has no real solution, temperature is a monotonic function of mass, otherwise is non-monotonic and isothermal solutions exist. Since the ADM mass $M$ \eqref{ADM_mass_general} is proportional to $m$ \eqref{mass}, and  the black hole temperature and mass $m$ are expressed in terms of the horizon radius in equations (\ref{mass}) and (\ref{temp}), it is preferred to study ${dm\over d\rho_H}$ and ${dT\over d\rho_H}$ rather than ${dT\over dM}$ directly. 
%
\begin{figure}
\begin{center}
\subfigure[Free energy]{{\includegraphics[scale=.75]{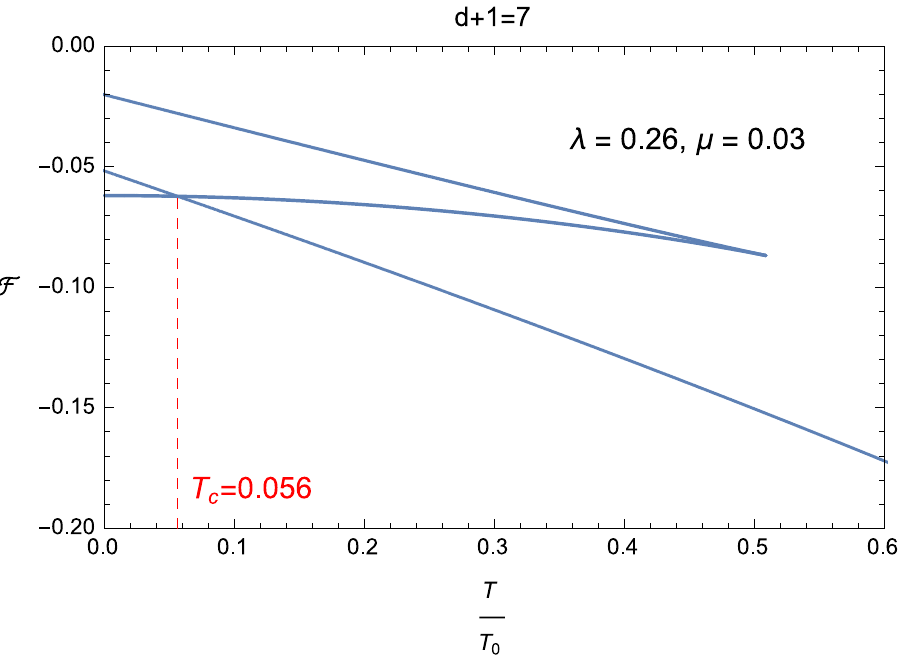} \label{fig:FvsTla26} }
\qquad{\includegraphics[scale=.75]{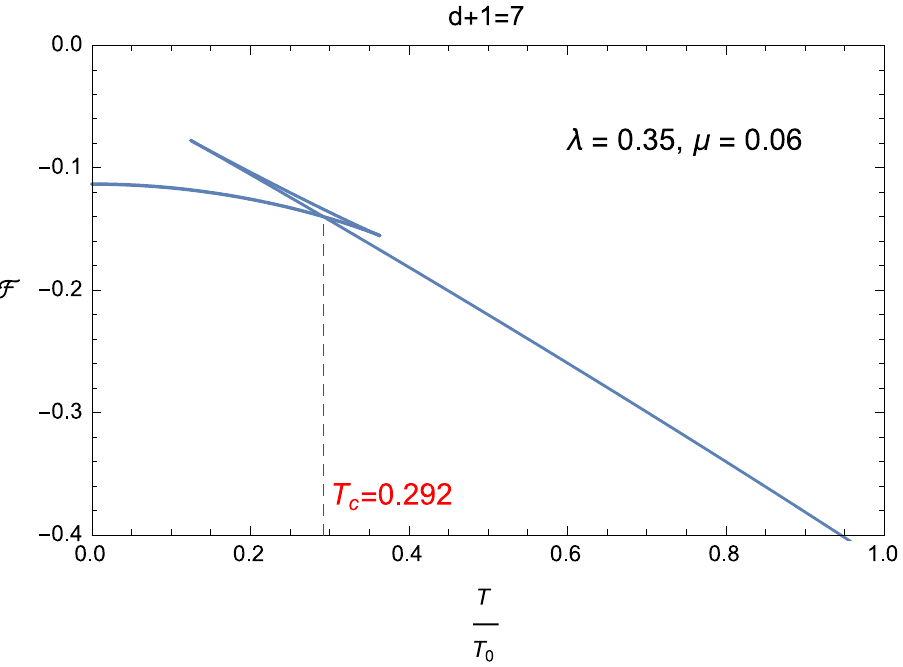} \label{fig:FvsTla35} }}
\subfigure[Entropy]{{ \includegraphics[scale=0.75]{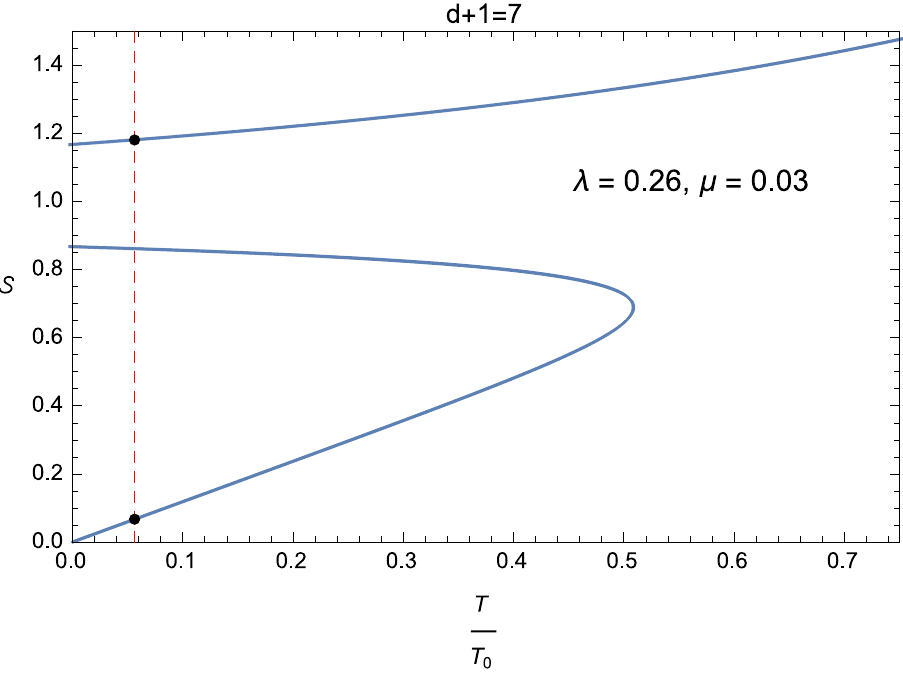}  \label{fig:SvsTla26}} 
\qquad
{ \includegraphics[scale=0.75]{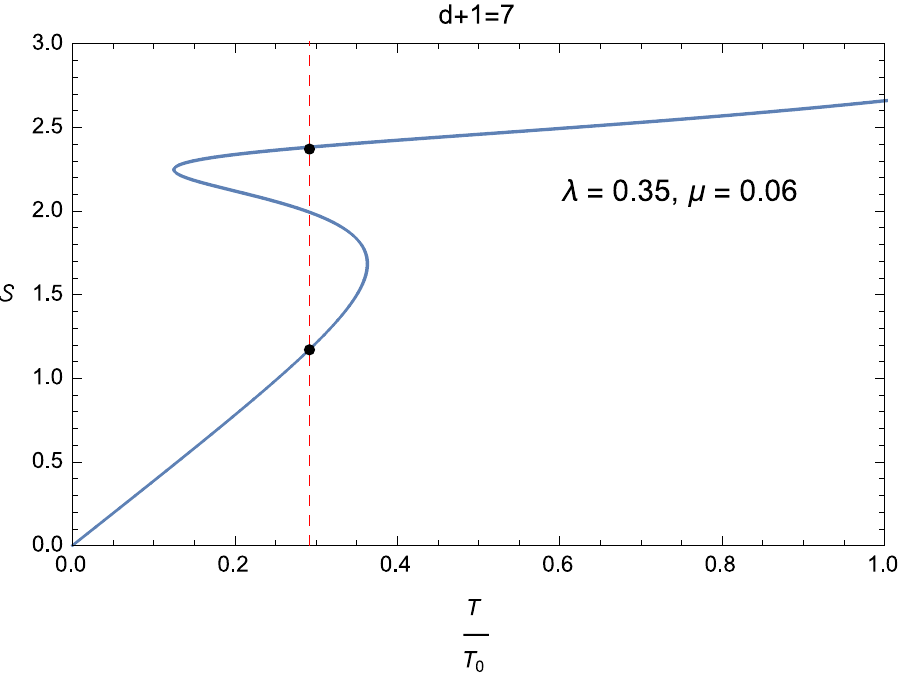}  \label{fig:SvsTla35}} }
\end{center}
\caption{(a) Free energy density $\mathcal F$ \eqref{curlyF} against temperature $T/T_0$ for a given $\{\la, \mu\}$ in 7-dimensions: on the left for $\la_c \le\la < \frac{1}{3}$ and $\mu>\frac{\la^2}{3}$ where a phase transition between two isothermal black holes with different masses occurs at $T_c \approx 0.056$; on the right for $1/3 < \la < \la_d=3/5$ and $\mu>\mu_{c_2}$ where a phase transition between two isothermal black holes with different masses occurs at $T_c \approx 0.292$. 
(b) Thermal entropy density $\mathcal S$ \eqref{curlyS} against temperature $T/T_0$ where the phase transition between two black holes in the lowest and highest branches are shown with black dots. The dashed red line indicates the temperature of phase transition and discontinuity in the thermal entropy reveals that the phase transition is of first order.} 
\end{figure}

In general, our analysis reveals that non-monotonicity of $T(M)$ is due to having either two extremal black holes or extrema in the temperature. 
The former is related to the behaviour of ${dm\over d\rho_H}$ (this simply follows from the thermodynamic relation ${dM\over d\rhoh}= T {dS\over d\rho}$), while for the latter one should inspect ${dT\over d\rho_H}$. 

For any $d > 6$, in order to find the solutions of the equation ${dm\over d\rho_H}=0$ at a non-trivial horizon radius $\rho_H\neq0$, we need to analyse a cubic equation in terms of $\rho_H^2$, {\it i.e.}
\begin{equation}\label{dmass}
d \rho _H^6-(d-2) \rho _H^4+(d-4) \lambda  \rho _H^2-(d-6) \mu=0\,. 
\end{equation}
For a given $\{\la, \mu\}$ if the discriminant of the cubic equation (\ref{dmass}) is positive, then ${dm\over d\rho_H}$ has three real roots which result in having two minima for the mass: $m_{ext_1}=m(\rho_{H<})$ and $m_{ext_2}=m(\rho_{H>})$ where $\rho_{H<}$ and $\rho_{H>}$ are the {\it smallest} and {\it largest} real roots of (\ref{dmass}), respectively. 
If $m_{ext_1} \le m_{ext_2}$ then there are two extremal black holes whose masses correspond to $m_{ext_{1, 2}}$. 
Note that if $m_{ext_1}  > m_{ext_2}$, then there is only one extremal black hole solution with $m_{ext_2}=m(\rho_{H>})$. In this case temperature is always a monotonic function of mass and no phase transition is expected. 

On the other hand, if the discriminant of equation (\ref{dmass}) for a given $\{\la, \mu\}$ is negative, ${dm \over d\rho_H}$ has only one real root and therefore there is only one extremal black hole. 
Nevertheless, in such a case $T(m)$ could still be non-monotonic  due to having more than one extremum: our investigations show that
equation ${dT\over d\rho_H}=0$ could have two (non-trivial) real distinct solutions. 
Hence, for a given $\{\la, \mu\}$ one needs to look for two real solutions of
\begin{eqnarray}\label{dtemp}
d \rhoh^{10} +  (d - 2 - 6 d \lambda ) \rhoh^8 +  (15 d \mu -(d-8) \lambda ) \rhoh^6 - 2  \left( 2 (d+3) \mu-(d-4) \lambda ^2 \right) \rhoh^4 \nonumber\\\hspace{.5cm}- 3 (d-8) \lambda  \mu  \rhoh^2 +3 (d-6) \mu ^2 = 0\,.
\end{eqnarray}
The equation (\ref{dtemp}) should be solved numerically in arbitrary dimensions, except for $d=6$ where one can find solutions analytically. 

The above analysis also applies to the 7-dimensional case ($d=6$). 
However, in this case equation (\ref{dmass}) is independent of $\mu$. 
Therefore, the behaviour of ${dm\over d\rho_H}$ for a given $\la$  is valid for all $\mu$'s, and here the only constraint on $\mu$ is that of the reality constraint, {\it i.e.} $\mu\ge\la^2/3$.
Instead, the behaviour of ${dT\over d\rho_H}$ still depends on both parameters $\{\la, \mu\}$. 

To summarize, for a given $\la\, (d\ge6)$ and $\mu\, (d>6)$ in order to specify non-monotonicity of $T(m)$ due to having two extremal black holes not only the discriminant of (\ref{dmass}) should be positive but also $m_{ext_1} \le m_{ext_2}$. Alternatively, for a given $\{\la, \mu\}$ in any  dimension $T(m)$ could be non-monotonic as $dT/d\rho_H=0$ might have more than one real solution.

\vskip 0.5 cm
To proceed further it is beneficial if we classify regions of $\la$ as below:

{\bf I)} $\la <  \la_c$: where $\la_c$ is obtained by solving equation (\ref{dmass}) for $\mu=\la_c^2/3$ while demanding $m_{ext_1}=m_{ext_2}$.\footnote{Some examples of $\lambda_c$ are: $\lambda_c=0.25$ in 7-dimensions ($d=6$), $\lambda_c=0.301836$ in 8-dimensions ($d=7$), $\lambda_c=0.316987$ in 9-dimensions ($d=8$) and $\lambda_c=0.323678$ in 10-dimensions ($d=9$).}
For any $\la < \la_c$ and $\mu \ge \la^2/3$ there is only one extremal black hole with $m=m_{ext_2}$ and horizon radius at $\rho_{H>}$ which is the largest real root of the equation (\ref{dmass}). Temperature is a monotonically increasing function of mass and therefore, no phase transition is expected in this range.

{\bf II)} $\la_c \le \la < 1/3$: in this region for any $\la^2/3 \le \mu \le \mu_{c_3}$ the discriminant of equation (\ref{dmass}) is positive and $m_{ext_1} \le m_{ext_2}$, therefore the system has two extremal black holes and $T(m)$ is non-monotonic. For a given $\la$ one can easily obtain $\mu_{c_3}$ by solving equation (\ref{dmass}) while demanding $m_{ext_1}=m_{ext_2}$. Note that $\mu_{c_3}\rightarrow\infty$ in 7-dimensions since equation (\ref{dmass}) is independent of $\mu$. 
Thus, in this range of $\{\la, \mu\}$ it is legitimate to expect a phase transition between smaller and larger isothermal black holes at some critical temperature $T_c \ge 0$. In order to check whether the phase transition happens or not, one should compare the free energy of the black hole solutions against temperature, {\it i.e.} $F(T)$. 
On the left panel, figure \ref{fig:FvsTla26} shows such an example where the occurrence of a phase transition is vivid at $T_c \approx 0.056$. 
Moreover, by examining black hole entropy $S(T)$ one finds that the phase transition is of first order, see figure \ref{fig:SvsTla26} on the left panel. 
Note that for $\la=\la_c$ or $\mu=\mu_{c_3}$ a phase transition happens at $T_c=0$, whereas $T_c$ increases by increasing $\la$ (or decreasing $\mu$) keeping fixed $\mu$ (or fixed $\la$). 
Furthermore, as the number of spacetime dimensions becomes larger, $\la_c$ approaches $1/3$.
\begin{figure}[t!]
\begin{center}
{{\includegraphics[scale=.59]{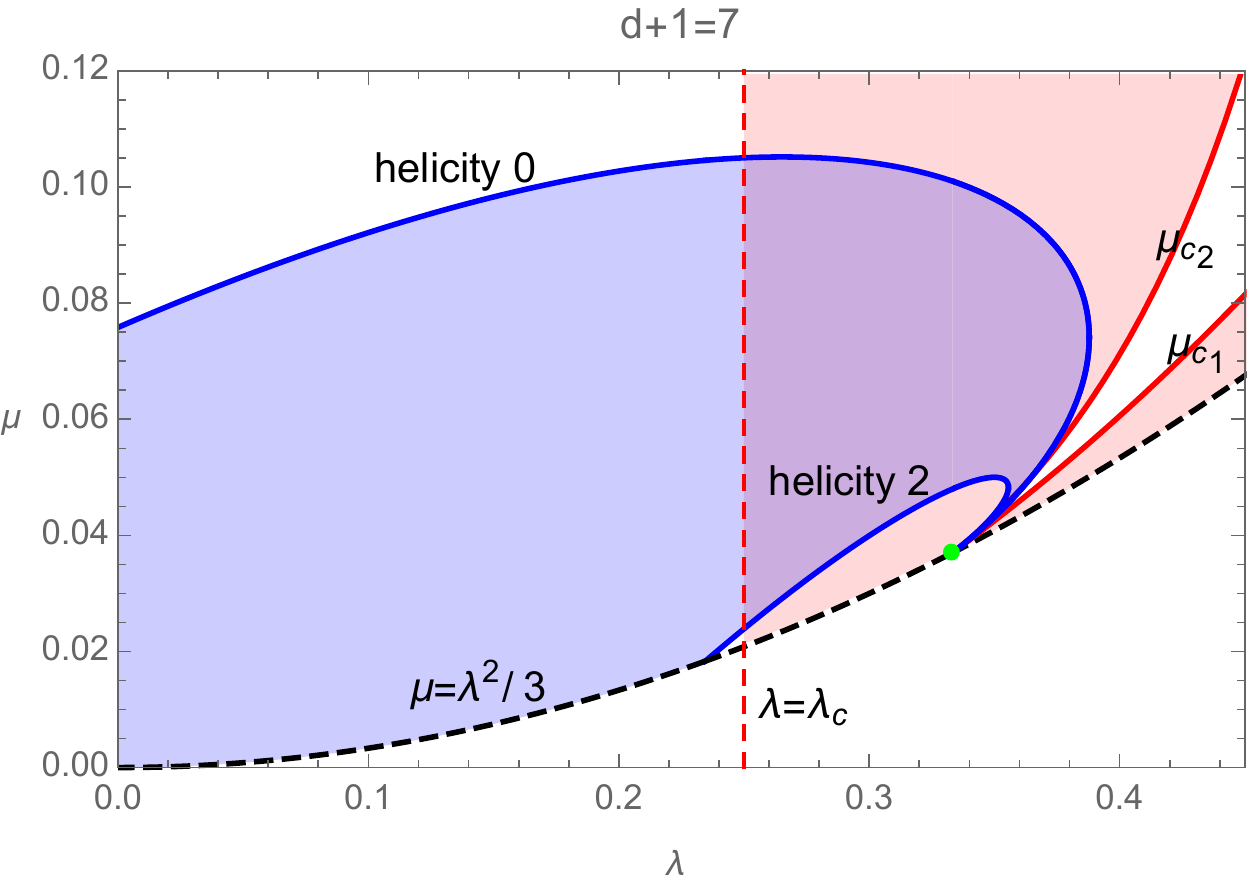} }
\,
{ \includegraphics[scale=.595]{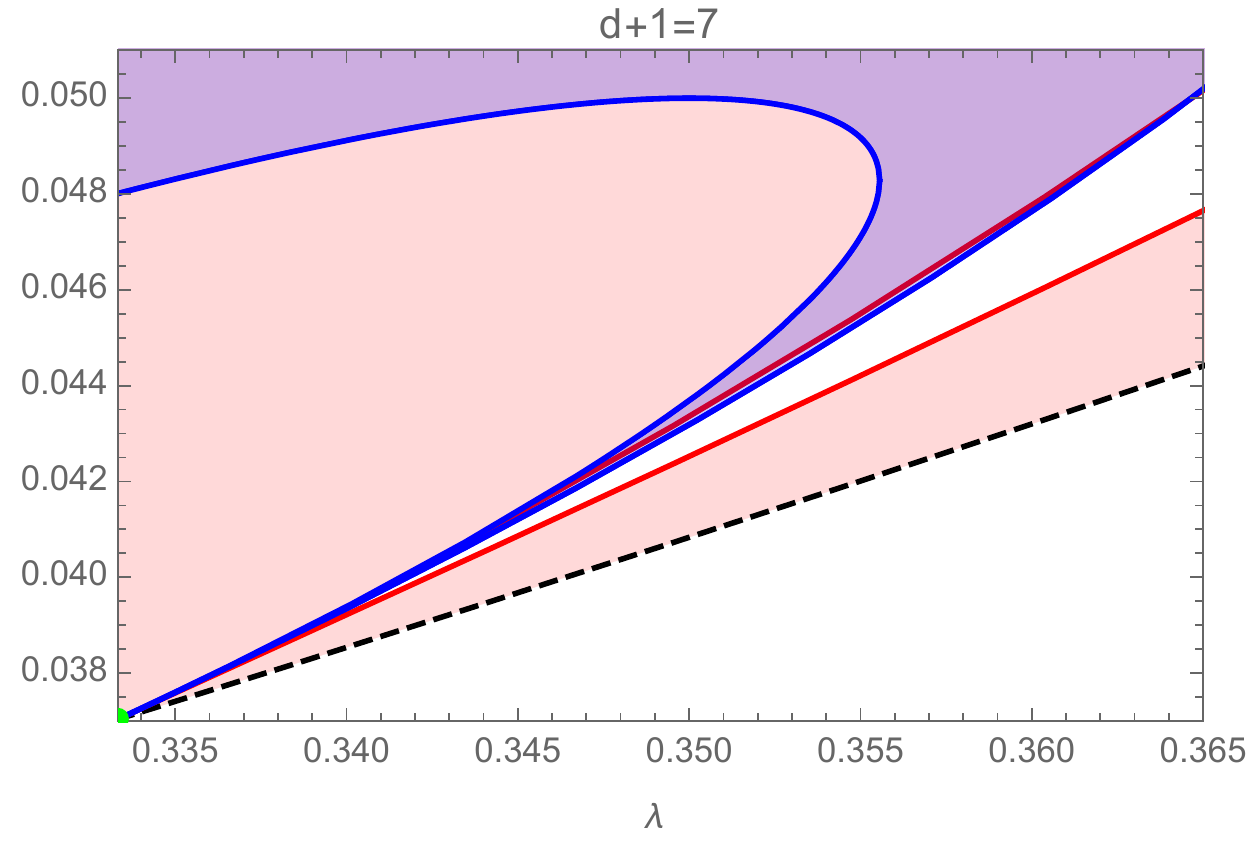}}}
\,
{{\includegraphics[scale=.59]{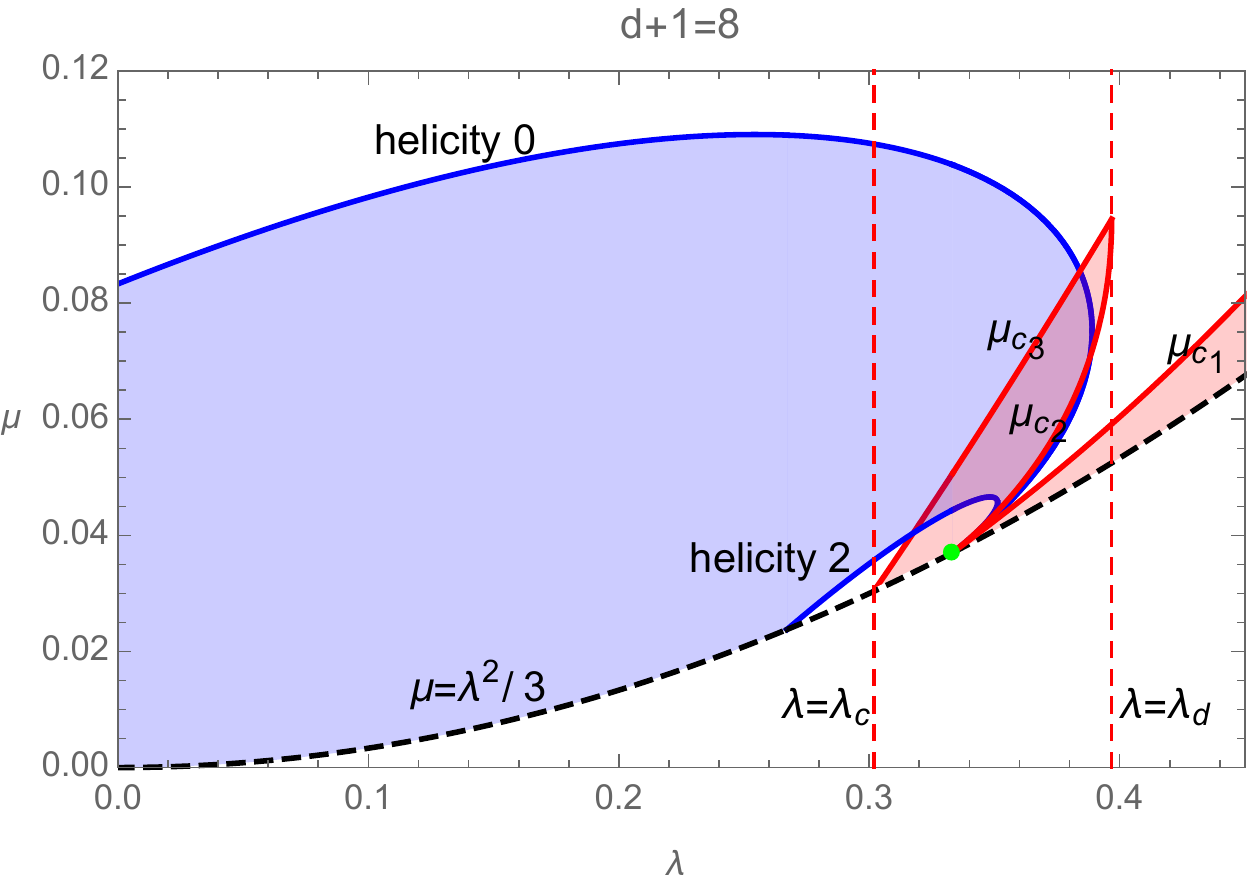} }
\,
{ \includegraphics[scale=.595]{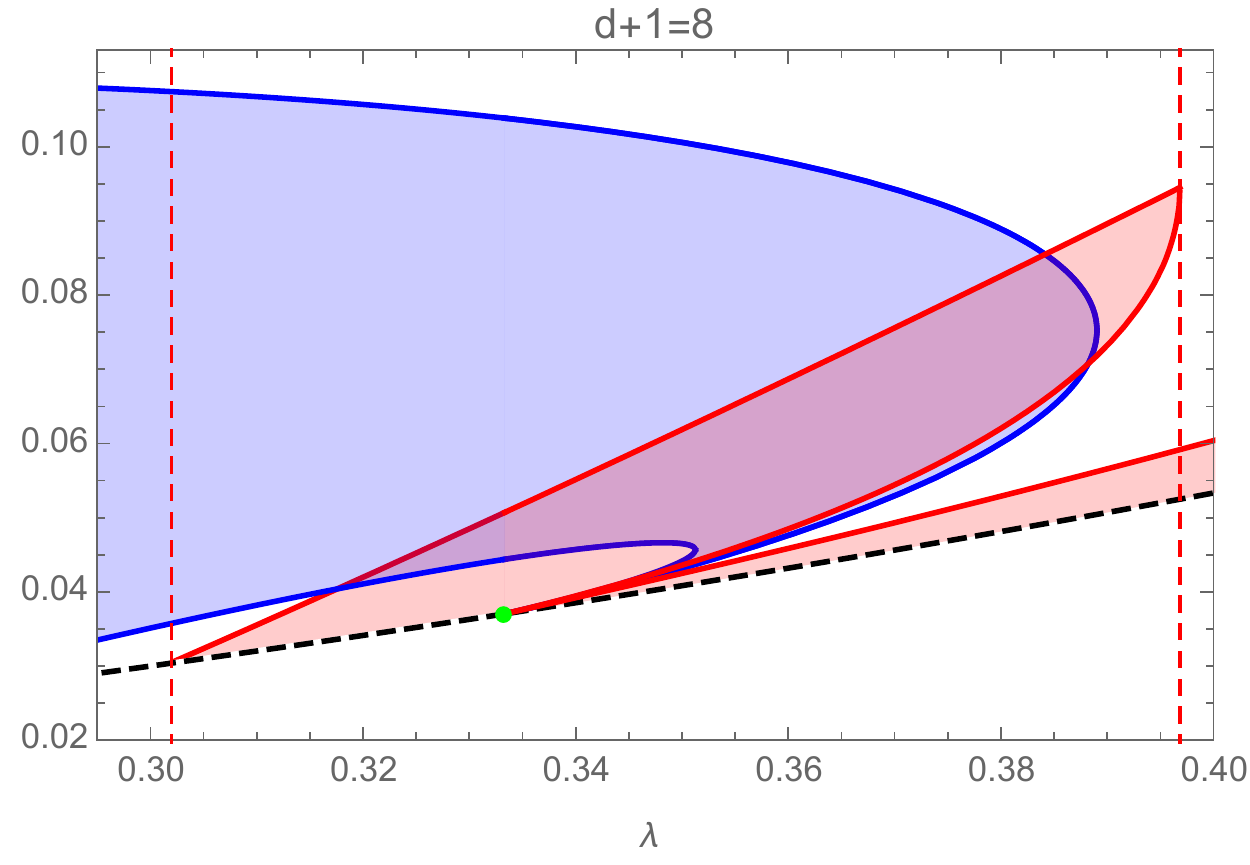}}}
\,
{{\includegraphics[scale=.59]{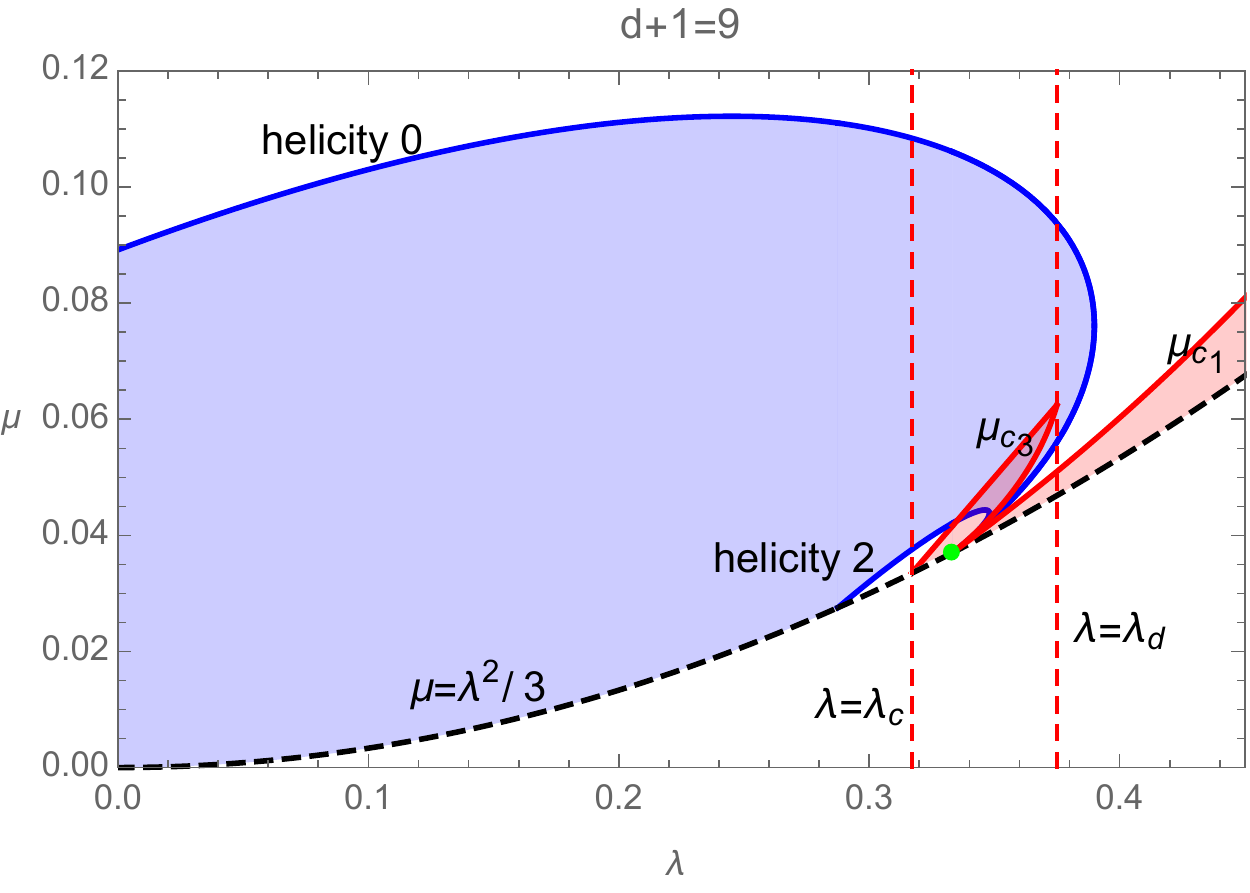} }
\,
{ \includegraphics[scale=.595]{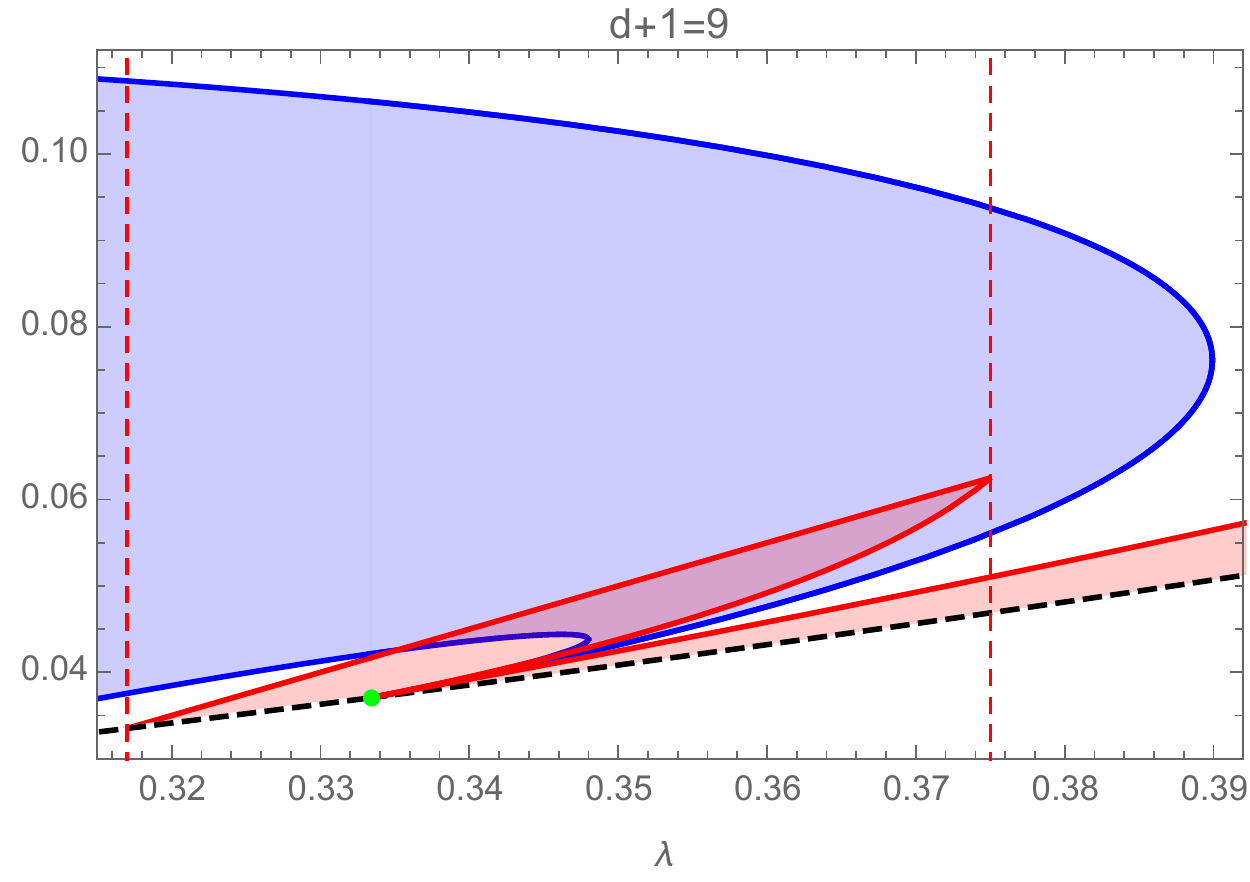}}}
\end{center}
\vspace{-1cm}
\caption{Parameter space of third order Lovelock in 7, 8 and 9-dimensions. The shadowed blue region is the allowed region due to boundary causality. The shadowed red region indicates the region where phase transitions between small and large black holes occur. The green dot represents the maximally symmetric AdS spacetime with $(\la, \mu)=(1/3, 1/27)$ for which no phase transition occurs. It is visible that as the spacetime dimensions increase the regions that phase transitions occur shrink. Also both $\la_c, \la_d\rightarrow 1/3$ when $d\rightarrow\infty$.   \label{fig:mulambda678}  }
\end{figure}

{\bf III)} $1/3 \le \lambda < \la_d$: where $\la_d=3/5$ in 7-dimensions ($d=6$) and $\la_d = (d - 2)^2/3 d (d - 4)$ in any higher dimension ($d>6$). Depending on the value of $\mu$ there might exist isothermal black holes due to having either an extremum in the temperature or two extremal black holes. 
Therefore, one should inspect non-monotonicity of temperature by examining both ${dT\over d\rho_H}$ and ${dm\over d\rho_H}$. 
Our analysis indicates that $T(m)$ is non-monotonic in two intervals: $\la^2/3 \le \mu \le \mu_{c_1}$ and $\mu_{c_2} \le \mu \le \mu_{c_3}$. For a given $\la$ both $\mu_{c_1}$ and $\mu_{c_2}$ are obtained by demanding that two real non-trivial solutions of equation (\ref{dtemp}) coincide. Whereas $\mu_{c_3}$ for $d>6$ is obtained by solving equation (\ref{dmass}) while demanding $m_{ext_1}=m_{ext_2}$. In 7-dimensions ($d=6$) there is no upper bound in the second interval since equation (\ref{dmass}) is independent of $\mu$.
As a result, non-monotonicity of temperature and possible phase transitions are expected in these two intervals. 
Again, in order to check the actual occurrence of phase transitions, one has to compare the free energy of coexisting solutions. 
An example of such comparison is shown on the right in figure \ref{fig:FvsTla35} in 7-dimensions, while the discontinuity of entropy indicates a first order phase transition on the right in figure \ref{fig:SvsTla35}. 
It is straightforward to work out similar comparisons of free energy in arbitrary dimensions to see that phase transitions between smaller and larger black holes always happen in the range where temperature is non-monotonic. 

The point $(\la, \mu)=(1/3, 1/27)$ is an exception in this range as the symmetry enhances  the spacetime to a full AdS space for which no phase transition happens.

{\bf IV)} $\la \ge \la_d$: in this region the discriminant of equation $(\ref{dmass})$ is strictly negative which results in having only one extremal black hole for all values of $\mu$. 
Therefore, the temperature could only be non-monotonic due to having an extremum which happens if $\la^2/3 \le \mu \le \mu_{c_1}$. Again $\mu_{c_1}$ is obtained by demanding that two real non-trivial solutions of equation (\ref{dtemp}) coincide. 
Then, we expect a possible phase transition in this range and comparing the free energy of isothermal black holes confirms the occurrence of a phase transition, which is first order as other regions. 
In dimension less than 9, $\la_d$ is larger than the maximum $\la$ allowed by causality. 
Furthermore, $\la_d$ approaches to $1/3$ as $d$ increases. 

Figure \ref{fig:mulambda678} is a complete parameter space in 7, 8 and 9-dimensions and shows a summary of possible phase transitions in all the regions discussed above. 
The blue regions consist of those values allowed by the boundary causality which was discussed earlier in section \ref{sec:CausCos}. The shadowed red region indicates the existence of phase transitions either due to having two extremal black holes or extrema in temperature. 
The green dot located at $(\la, \mu)=(1/3, 1/27)$ represents the maximally symmetric AdS space for which no phase transition happens. 
Moving from 7 to 8-dimensions, the size of the regions where phase transitions happen dramatically reduces.
This is due to the fact that the upper limit $\mu_{c_3}$ is absent in 7-dimension, since equation (\ref{dmass}) is independent of $\mu$ for $d=6$, as we explained above. 
From figure \ref{fig:mulambda678} it is also evident that as dimensionality increases, the red areas in the shaded blue regions (that is allowed by causality) shrink and eventually disappear as $d\rightarrow\infty$ since in this limit both $\la_c, \la_d$ approach to $1/3$.
This means that we approach to the green dot in parameter space as $d\rightarrow\infty$, and no phase transition happens at this point. 
In another words, the theory is stable for a wider range of Lovelock parameters in higher dimensions.

\section{Results and Discussion}
\label{sec:RE-Lovelock}
\begin{figure}
\begin{center}
\includegraphics[scale=1]{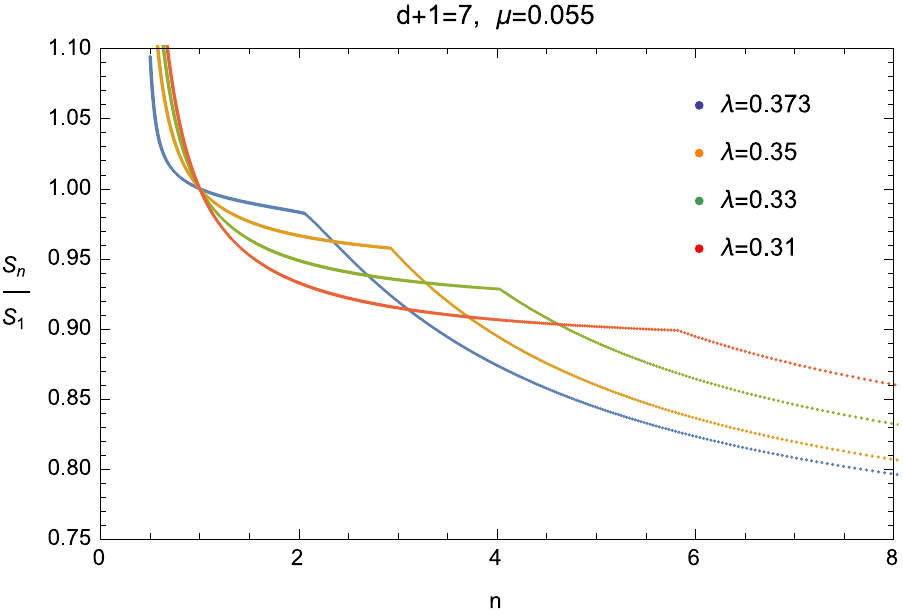}
\caption{Ratio of R\'enyi entropy to the entanglement entropy as a function of $n$ in 7-dimensions for a fixed $\mu=0.055$ and $\la=0.0373, 0.35, 0.33, 0.31$ from top to bottom. Lovelock parameters have been chosen from the causal region in parameter space where the phase transition in the bulk occur, figure \ref{fig:mulambda678}. Therefore, each curve displays a kink in the R\'enyi entropy at $n_c=T_0/T_c$. The kink appears at larger $n_c$ for smaller $\la$ and moves to the left towards $n=1$ as $\la$ grows. Note that for $\mu=0.055$ an upper bound from boundary causality, {\it i.e.} $\la \le 0.373$, imposes a lower bound on $n_c\ge 2.06$.\label{fig:Snmu055} }
\end{center}
\end{figure}

We can now investigate the implications of the instabilities of third order Lovelock black holes studied in section \ref{sec:phase-trans} on the R\'enyi entropies.   
As explained in section \ref{sec:RE-HRE}, in order to calculate the R\'enyi entropy of a boundary CFT, we can either use the expression \eqref{RenyiEnt},
where now $S_{thermal}$ is the black hole thermal entropy in the bulk as a function of its temperature given by expressions \eqref{entropy_general_d} and \eqref{temp}, or we can use \eqref{SnFT} where now $F$ is the black hole free energy \eqref{def_free_en} again as a function of the temperature \eqref{temp}. 

Let us consider for example the expression \eqref{RenyiEnt}. Recall that $T_0$ is the temperature of the boundary CFT, and we use this value as a reference temperature, while the final temperature is given by $T_0/n$. 
Keeping fixed $T_0$,  whenever the final temperature is smaller than $T_c$,  we end up integrating over a piece-wise continuous function  $S_{thermal}$. 
This becomes clear by looking at figure \ref{fig:SvsTla26}. 
Thus, the integral over the jump between the two stable branches will result in a continuous but not differentiable function of ${T\over T_0}$,  and this is nothing but the R\'enyi entropy, cf. \eqref{RenyiEnt}. 
Figure \ref{fig:Snmu055} shows 
the R\'enyi entropy in terms of the index $n(={T_0\over T})$ in 7-dimensions for a fixed $\mu=0.055$ and several $\la$'s, all in the causal region where the phase transition happens in the bulk. 
It is evident that there is a kink in the R\'enyi entropy at $n_c={T_0\over T_c}$ which is a direct consequence of the bulk first order phase transition. 
In particular, in any dimension the kink is placed at $n_c>1$.
This is due to the fact that the phase transitions in the causal regions occur at a critical temperature that is always smaller than $T_0$, {\it i.e.} ${T_c\over T_0}<1$. 
As expected from field theoretical computations, R\'enyi entropies are divergent when the index $n$ approaches to zero (in terms of the entanglement spectrum this limit represents the logarithm of the number of non-vanishing eigenvalues), specifically the leading divergence behaves as ${1\over n^{d-1}}$. 
On the other side, they approach  a constant as $n\to\infty$ (which is proportional to the logarithm of the largest eigenvalue), where again the specific value of the constant depends on the dimension $d$ and the coupling constant $\{\la, \mu\}$, see for example the discussion in \cite{Hung:2011nu}.

\begin{figure}
\begin{center}
\includegraphics[scale=.9]{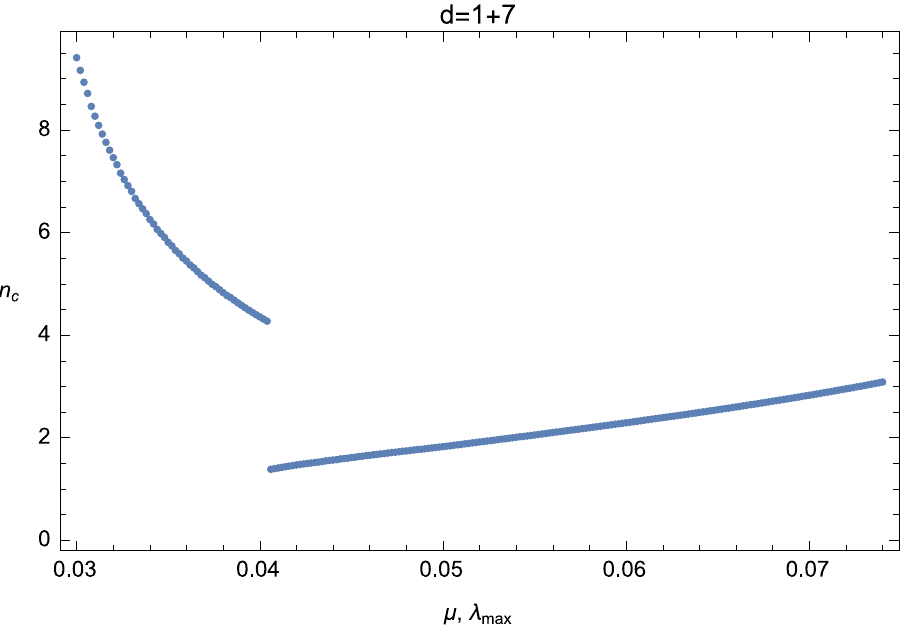}
\caption{$n_{c_{min}}$ for $\mu$ and corresponding maximum value of $\la$ allowed by causality for which phase transition occurs. The discontinuity in $n_c$ at $\mu=0.0406$ is due to the jump in $\la_{max}$ from the helicity 2 curve to the curve $\mu_{c_2}$ in figure \ref{fig:mulambda678}. 
The lowest value of $n_{c_{min}}=1.38$ is at $\mu=0.0406$ for that $\la_{max}=0.343$.\label{fig:nc} }
\end{center}
\end{figure}

In general, in a given dimension for a fixed $\mu$, decreasing (increasing) $\la$ leads to an increase (decrease) in $n_c$. 
There is always a lower bound on $\la$ given by $\la_c$ for which $n_c\rightarrow\infty$: recall that for $\la=\la_c$ the phase transition happens at ${T_c\over T_0}=0$ (cf. region II in section \ref{sec:phase-trans}). 
Moreover, $\la\le\la_{max}$ (for a fixed $\mu$) which imposes a lower bound on $n_c$, namely $n_{c_{min}}$: depending on the value of $\mu$ the upper limit $\la_{max}$ is either dictated by causality constraints (\ref{CausCons}) or is the maximum possible $\la$ in the causal region for which a phase transition happens. 
Figure \ref{fig:nc} shows $n_{c_{min}}$ in 7-dimensions for $0.03\le\mu\le 0.074$ and the corresponding $\la_{max}$ which is partly obtained by causality constraints: helicity 2 in the region $0.03\le\mu\le0.0405$ and helicity 0 for $0.0499\le\mu\le 0.0741$.  However, in the range $0.0405<\mu<0.0499$, $\la_{max}$ is the maximum value in the causal region for which phase transitions happen and it belongs to the curve $\mu_{c_2}$ in figure \ref{fig:mulambda678} for 7-dimensions. 
Notice that there is a discontinuity in $n_c$ at $\mu=0.0406$ due to the jump in $\la_{max}$ from the helicity 2 curve to the $\mu_{c_2}$ curve in figure \ref{fig:mulambda678}.
Also in figure \ref{fig:nc} the lowest value of $n_{c_{min}}=1.38$ (for $\mu=0.0406$ and $\la_{max}=0.343$) indicates that for any fixed $\mu$ in 7-dimensions $n_c\ge 1.38$, {\it i.e.} the kink does not happen at or very close to 1. 
Therefore,  despite having a kink the R\'enyi entropy is still smooth and differentiable in the vicinity of $n=1$. 

Alternatively, in a given dimension one can examine how $n_c$ varies with $\mu$ for a fixed $\la$: our analysis reveals that $n_c$ increases (decreases) with increasing (decreasing) $\mu$. 
Therefore, for each $\la$ the minimum value of $n_c$, namely $n_{c_{min}}$, is obtained for the minimum $\mu$ in the causal region for which the phase transition happens. 
From figure \ref{fig:mulambda678} it is easy to see that in 7-dimensions $\mu_{min}$ is partly obtained by causality constraints,
in particular by the helicity 2 constraint for $\la_c \le \la \le 0.343$ and by the helicity 0 constraint for $0.364 \le \la < 0.389$. 
Whereas $\mu_{min}$ belongs to the curve $\mu_{c_2}$ if $0.343 < \la < 0.364$. One can reproduce a plot similar to \ref{fig:nc} for $n_{c_{min}}$ but at a fixed $\la$ and correspondent $\mu_{min}$. 
However, the lowest value of $n_c$ is still $n_{c_{min}}=1.38$, obtained for $\la=0.343$ $\mu_{min}=0.0406$.
Hence, we have again that $n_c \ge 1.38$ and the kink is far enough from $n=1$ to give a differentiable R\'enyi entropy in the vicinity of $n=1$.

\vskip 0.5 cm

To our knowledge this is the first pure gravitational system to produce first order phase transitions in the bulk which are reflected in a kink of the dual R\'enyi entropy. 
In holographic field theories, second order phase transitions were previously  discussed in \cite{Belin:2013dva, Belin:2014mva}.%
\footnote{For related discussions in Gauss-Bonnet gravity see also \cite{Pastras:2014oka, Pastras:2015mza}.}
There, the second derivative of the R\'enyi entropies with respect to the index $n$ was found to be discontinuous. 
We stress that there the bulk mechanism to give arise the phase transition is rather different: it is either due to the formation of hairy black holes in presence of light scalars \cite{Belin:2013dva} or due to a holographic superconductor-like mechanism in the charged case \cite{Belin:2014mva}. 
Our original boundary field theory is a CFT at zero temperature living in a flat $d$-dimensional space where we have a bipartite system separated by a $(d-2)$-dimensional spherical entangling surface of radius $R$. 
The only scale present here is set by the radius of the entangling surface. 
We are essentially probing the ground state of this bipartite system. 
Our results suggest that there is an emergent critical index $n_c$ (where the R\'enyi entropy displays a kink) which might be a sign of a phase transition in the ground state: that is the spectrum seems to have distinct regions, likely characterised by two distinct probability distributions. 

An analogous non-analytic dependence was found  in the universal coefficients of R\'enyi entropy for the $\mathrm{O}(N)$ model close to critical points \cite{Metlitski:2009iyg} (both in the large $N$-limit and $4-\epsilon$-expansion).%
\footnote{We want to stress that in this work we are not extracting the universal coefficients, we are computing the whole value of R\'enyi entropy.}
This was found by purely field theoretic considerations but  $\mathrm{O}(N)$ vector models are conjectured to be dual to higher spin theory in AdS \cite{Klebanov:aa, Sezgin:aa},%
\footnote{Cf. the recent review \cite{Giombi:2016aa} and references therein.}
suggesting that similar behaviours to that found in our study can also been seen in another gravitational setting. 
Another field theoretical example is provided by the work~\cite{Stephan:2011aa}.
Here, the authors find a phase transition in the  R\'enyi entropy for Luttinger liquids at a critical $n_c$, which emerges essentially when the index $n$ has a significant effect on the natural scale of the field theory (Luttinger parameter).
An important lesson from \cite{Stephan:2011aa} is that the replica method would miss such a phase transition, and a general caution should be kept in mind in applying the replica method in cases where R\'enyi entropy is not analytical. 
Nevertheless, as already pointed out in \cite{Belin:2014mva} the fact that R\'enyi entropy might not be analytical does not have any effect in the proof of RT formula where only analyticity at $n=1$ is assumed~\cite{Lewkowycz:2013nqa, Dong:2016hjy}.  

As mentioned at the beginning, this holographic set-up could provide a simpler and yet rich framework where novel aspects of strongly coupled higher dimensional CFT's could be revealed.
It would be interesting to investigate how the inclusion of a $\mathrm{U}(1)$ charge in our model would affect the bulk instabilities, and thus the phase transitions in R\'enyi entropy.
A major challenge would be how to holographically realises the non-analytic behaviour of the R\'enyi entropies found in \cite{Metlitski:2009iyg}, we leave this for future works.

\section*{Acknowledgements}

We acknowledge useful discussions with F. Bigazzi, J. Camps, P. Di Vecchia, S. Hirano, C. Keeler, M. Kulaxizi, P. Pani, SJ. Rey, S. Sachdev, M. Smolkin, M. Taylor. 
We are indebted to R. Myers and L. Thorlacius  for many enlightening comments and discussions.
We thank L. Thorlacius for reviewing the manuscript. 
VGMP would like to thank NORDITA for hospitality during the initial part of this work. 
RP thanks the Perimeter Institute for Theoretical Physics for hospitality during the completion of this work.
This research was supported in part by the Icelandic Research Fund under contracts 163419-051 and 163422-051, and by grants from the University of Iceland Research Fund.

\bibliographystyle{JHEP}
\bibliography{RHL}

\end{document}